\documentclass[aps,pra,twocolumn,showpacs,superscriptaddress,floatfix]{revtex4-2}

\usepackage{graphicx}
\usepackage{dcolumn}
\usepackage{bm}
\usepackage{hyperref}
\usepackage{braket}
\usepackage{footmisc}
\usepackage[dvipsnames]{xcolor}
\usepackage{comment}
\usepackage{amsmath}
\usepackage{amssymb}
\usepackage{bm}
\usepackage{revsymb}
\usepackage{verbatim}
\usepackage{multirow}
\usepackage[normalem]{ulem}

\begin{document}

\preprint{APS/123-QED}

\title{Hyperfine and Zeeman Optical Pumping and Transverse Laser Cooling of a Thermal Atomic Beam of Dysprosium Using a Single 421\,nm Laser}


\author{Rohan Chakravarthy}
\email{Present address rohan.chakravarthy@ptb.de}
\affiliation{Johannes Gutenberg-Universit\"at Mainz, 55128 Mainz, Germany}
\affiliation{Helmholtz-Institut Mainz, GSI Helmholtzzentrum f{\"u}r Schwerionenforschung, 55128 Mainz, Germany}

\author{Jonathan Agil}
\affiliation{Johannes Gutenberg-Universit\"at Mainz, 55128 Mainz, Germany}
\affiliation{Helmholtz-Institut Mainz, GSI Helmholtzzentrum f{\"u}r Schwerionenforschung, 55128 Mainz, Germany}

\author{Arijit Sharma}
\affiliation{Department of Physics, Indian Institute of Technology Tirupati, Yerpedu 517619, Andhra Pradesh, India}
\affiliation{Center for Atomic, Molecular and Optical Sciences $\&$ Technologies, Indian Institute of Technology Tirupati, Yerpedu 517619, Andhra Pradesh, India}

\author{Jung Bog Kim}
\email{jbkim@knue.ac.kr}
\affiliation{Korea National University of Education, 250 Taeseongtabyeon-ro, Cheongju, South Korea}
\affiliation{Helmholtz-Institut Mainz, GSI Helmholtzzentrum f{\"u}r Schwerionenforschung, 55128 Mainz, Germany}

\author{Dmitry Budker}
\email{budker@uni-mainz.de}
\affiliation{Johannes Gutenberg-Universit\"at Mainz, 55128 Mainz, Germany}
\affiliation{Helmholtz-Institut Mainz, GSI Helmholtzzentrum f{\"u}r Schwerionenforschung, 55128 Mainz, Germany}
\affiliation{Department of Physics, University of California, Berkeley, CA 94720, USA}


\begin{abstract}

We demonstrate the effect of Zeeman and hyperfine optical pumping and transverse laser cooling of a dysprosium (Dy) atomic beam on the $4f^{10}6s^2(J=8) \rightarrow 4f^{10}6s6p(J=9)$ transition at 421.291\,nm. For $^{163}$Dy, a custom built electro-optic modulator is used to generate five frequency sidebands required to pump the atoms to the $F=10.5$ ground state hyperfine level and the light polarization is chosen to pump the atoms to the $m_F = 10.5$ Zeeman sublevel. The atoms are simultaneously laser-cooled using a standing wave orthogonal to the atomic beam. The resulting polarized and cooled atomic beam will be used in fundamental physics experiments taking advantage of the ``accidental'' degeneracy of excited states in Dy including the ongoing measurement of parity violation in this system.

\end{abstract}

\maketitle


\section{Introduction}

Dysprosium (Dy) is an atom of choice  in a wide variety of experiments due to its large ground-state magnetic moment (the largest currently known of all atoms), seven stable isotopes with mass numbers ranging from $A=156$ to 164, and the existence of a pair of metastable excited states of the same angular momentum and opposite parity that are nearly degenerate, so that complete degeneracy can be achieved via Zeeman tuning in a magnetic field as low as 1.4\,G\,\cite{Budker1994_ExpInvDy}. These nearly-degenerate levels were used in searches for temporal variation of the fine-structure constant $\alpha$\,\cite{Cingoez2007_alpha_var, Leefer2013_alpha_var} and its dependence on the gravitational potential\,\cite{Ferrel2007_gravitational_dependence_alpha}, tests of local Lorentz invariance and Einstein equivalence principle\,\cite{Hohensee2013_violations_lorentz_symmetry}, and a search for parity non-conservation\,\cite{Nguyen1997_PNC_Dy} (so far focused on $^{163}$Dy). It has also been used in searches for apparent oscillation of $\alpha$ induced by ultralight scalar dark matter\,\cite{Tilburg2015_ULDM, Xue2023_UDarkMatter}. 

Dysprosium atoms can be laser cooled\,\cite{Muehlbauer2018_optimization_cooling_Dy, Leefer2010_Transverse_cooling, Peterson2019_sawtooth_slowing_Dy}, trapped\,\cite{Youn2010_Dy_MOT, Bloch2023_Dy_tweezer_array} and cooled to quantum degeneracy\,\cite{Lu2011_BEC_Dy, Lu2012_fermi_gas}, and has become a favored system for the study of dipolar quantum gases\,\cite{Klaus2022_vortices_in_condensate, chomaz2022_dipolar_gases}.

Much of this work requires knowledge of the spectra of dysprosium, which was studied both theoretically\,\cite{Childs1970_ground_state_moments, Dzuba2010_states_Dy} and experimentally\,\cite{Studer2018_1001nmDy, Budker1991_Spectroscopy_Dy, Schmitt2013_narrow_line_transition_Dy} in the past few decades. Controlling the internal states of Dy is important for this work, including the ability to efficiently optically pump the atoms to a particular Zeeman sublevel\,\cite{Lecomte2025_production_of_stabilized_dipolar_gas} and to a particular hyperfine level in the case of the isotopes with nonzero nuclear spin.

In this work, we use a weakly collimated atomic beam with characteristic transverse velocities of $\simeq 20$\,m/s. A collimated narrow-band continuous wave laser only interacts with a small fraction of these atoms. Efficient population transfer of the atomic beam to an excited state requires either broadening the laser linewidth, matching the divergence of the atomic beam and the laser beam \cite{Nguyen2000_Dy_population_transfer}, or reducing the transverse velocities of the atomic beam by laser cooling. Laser cooling the atomic beam relies on the optical molasses technique, where a standing wave laser beam orthogonal to the atomic beam is used to reduce the transverse Doppler width. 

In this paper, we demonstrate optical pumping and laser cooling on an atomic beam of Dy, including the $^{163}$Dy isotope having well resolved hyperfine structure with energy differences $>1$\,GHz, with a single laser beam and a custom built electro optic modulator (EOM). We also demonstrate ``accidental'' optical pumping of the $^{161}$Dy isotope. The technique adopted in this work can facilitate state preparation in systems with complex hyperfine structure without the need for several repumper laser beams or the use of multiple EOM and acousto-optic modulator (AOM) systems. The results presented in this paper is an important step in improving the signal-to-noise ratio in the ongoing search for parity non-conservation (PNC) in Dy\,\cite{Leefer2014_towards_PNC}.


\section{Experimental Setup}

\begin{figure}[t]
    \centering
    \includegraphics[width=\columnwidth]{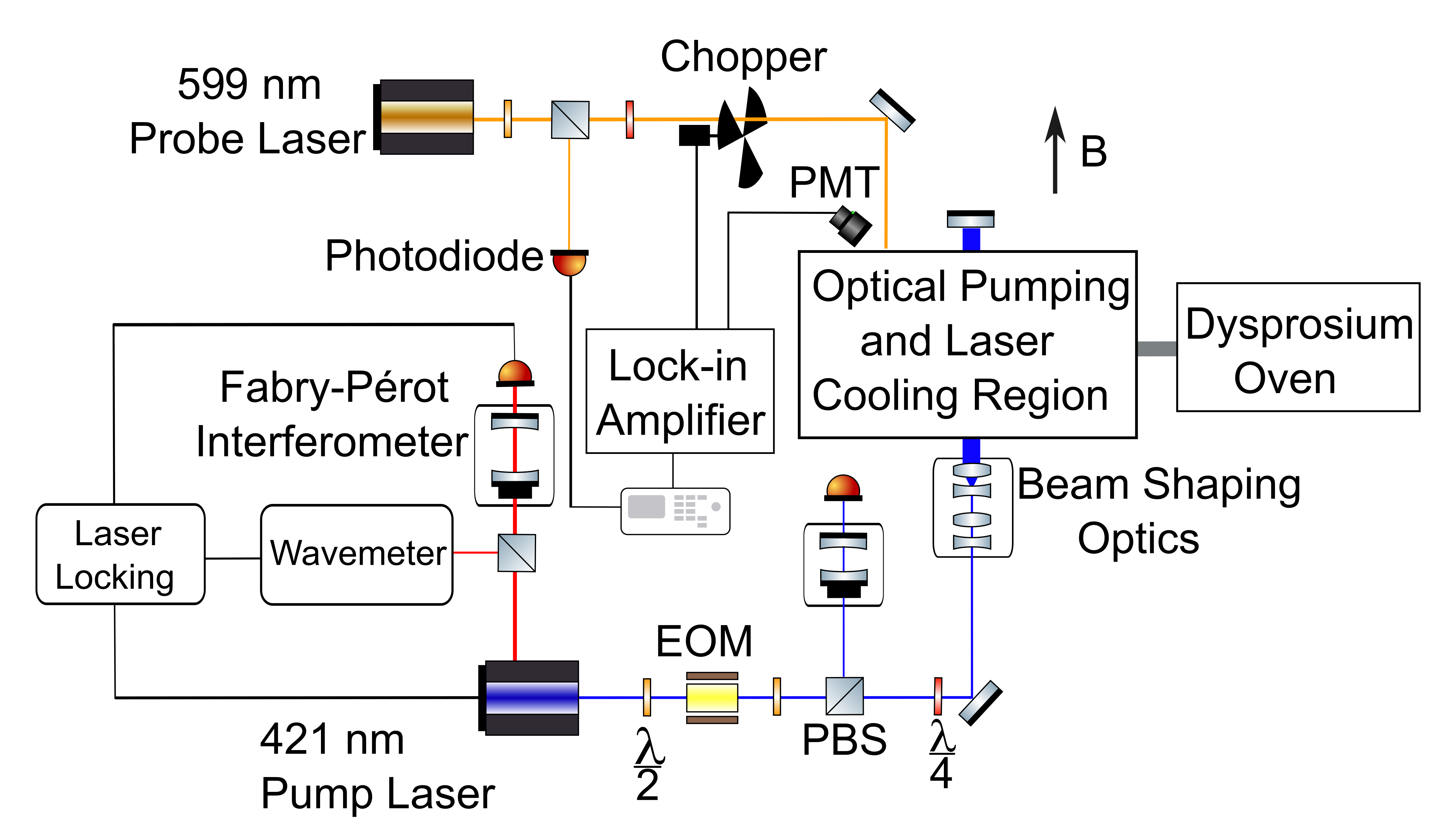}
    \caption{\label{fig:optsetup} Simplified schematic of the experimental setup. In the figure EOM is the Electro-Optic Modulator, PBS is the Polarising Beamsplitter, $\lambda/2$ and $\lambda/4$ are the half and quarter waveplates, PMT is the photomultiplier tube, UHV chamber is the ultra-high vacuum chamber, and B is the magnetic field defining the quantization axis. Of the two separate Fabry-Perot interferometers in the figure, one is used to lock the fundamental laser frequency of the 421\,nm laser and the second as a Fabry-Perot spectrum analyser for the second harmonic.} 
\end{figure}

\begin{figure}[b]
    \centering
    \includegraphics[width=\columnwidth]{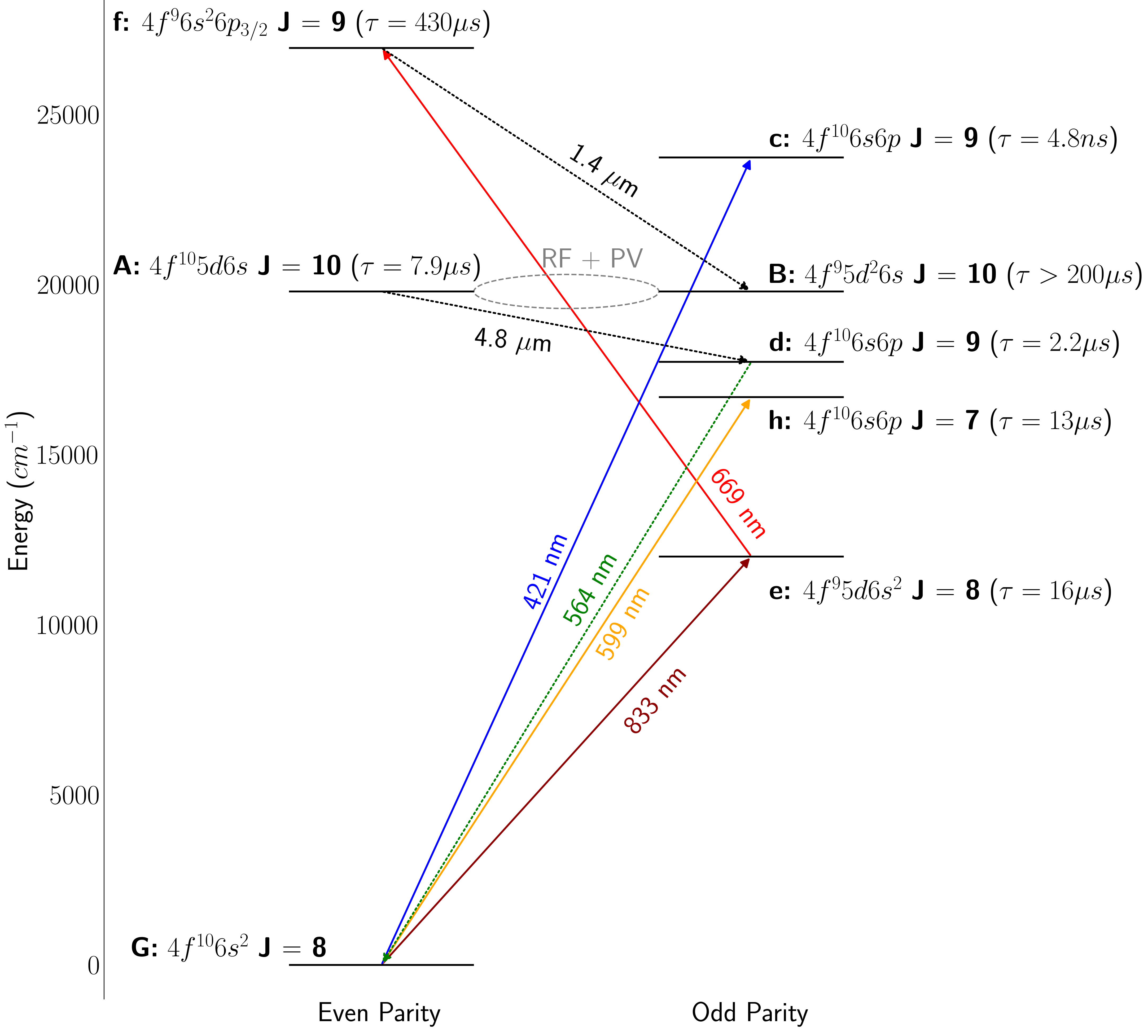}
    \caption{\label{fig:leveldiag} Diagram of the energy level of dysprosium representing the nearly degenerate pair of opposite parity states A and B. The diagram shows the transitions used in this work to pump and probe the ground state, respectively 421 and 599\,nm, as well as the transitions used for the B state preparation, 833 and 669\,nm, and the 564\,nm fluorescence detection of the parity signal. The A and B states are mixed with a RF field.}
\end{figure}

A schematic of the experimental setup is presented in Fig.\,\ref{fig:optsetup}. 
A sample of dysprosium is heated up to $\simeq 1500$\,$^\circ$C to generate the atomic beam. A more detailed description of the source and the UHV setup can be found in Ref.\,\cite{Weber2013_ac_stark_shifts_Dy}. Two pairs of coils installed inside the optical pumping region along with a pair of coils installed outside the experimental setup were used for the control of magnetic fields inside the optical pumping region. We use the strong 421\,nm $J=8\rightarrow J'=9$ transition\,\cite{Leefer2009_Hyperfine_structure_Dy} with a lifetime $\tau = 4.8$\,ns and a natural width $\Gamma/{2\pi} = 33$\,MHz\,\cite{Curry1997_Dy_radiative_lifetimes} for optical pumping and laser cooling, and a weaker 599\,nm $J=8\rightarrow J'=7$ transition\,\cite{Childs1977_598nm} ($\tau = 13$\,$\mu$s; $\Gamma/{2\pi} = 12$\,kHz)\cite{NIST_ASD} as a spectroscopic probe. A diagram representing the relevant energy levels for the measurement of optical pumping, laser cooling and PNC in Dy is represented in Fig.\,\ref{fig:leveldiag}.

The 421\,nm pump laser is a frequency-doubled master oscillator-power amplifier (MOPA) Toptica DLC-PRO system with an output power of $\simeq 400$\,mW. The laser beam is guided as shown in Fig.\,\ref{fig:optsetup} into the optical pumping region through a quarter wave plate and beam-shaping optics to obtain a rectangular beam with a width (along the direction of the atomic beam) of 20\,mm and height of 2\,mm leading to a typical interaction time of 65\,$\mu$s. The purpose of the rectangular beam shaping is twofold. First, the height is chosen so the laser covers a significant portion of the atomic beam. Second, the length allows for enough pumping cycles to occur ensuring full Zeeman pumping and cooling of the atomic beam ($\simeq 10^4$), see Appendix\,\ref{AppA}. The laser power delivered into the optical pumping/laser cooling region is $\simeq 230$\,mW. A part of the fundamental laser frequency output is guided into a Fabry-Perot cavity. The laser is locked to the cavity using a side-of-fringe lock and is then software locked to the wavemeter. The wavemeter is then used as a reference to tune the laser to the relevant Dy transition. A thirty minute measurement of the locked pump laser indicates deviations in frequency of $<0.5$\,MHz.

The probe is a frequency-doubled optical parametric oscillator (OPO) laser (Huebner C-Wave) with a linewidth of $\leq 1$\,MHz operating at 599\,nm with an output power of 50\,mW. After passing through a fiber, and polarization optics, a 1\,mm beam diameter laser beam with a power of 1\,mW is delivered to the experiment and $<2$\,mW of laser power to a reference photodiode. Both the pump and probe laser beams are circularly polarized to observe hyperfine and Zeeman pumping effects.

The probe laser is scanned across resonance to observe the spectrum of the 599\,nm transition of Dy while the pump laser is frequency-locked. Scanning the OPO laser results in fluctuations in the laser intensity delivered to the atoms and is monitored by the reference photodiode. The scanning probe laser drifts by $\simeq 100$\,MHz over a period of thirty minutes which leads to noticeable shifts over scans. The 599\,nm fluorescent light is detected using a photomultiplier tube (PMT, Hamamatsu R943-02) oriented orthogonal to the laser and atomic beam, with a 600\,nm central wavelength bandpass filter (10\,nm FWHM bandwidth, ThorLabs), a colour glass and a homemade spatial filter to reject scattered light from the optical viewports. We measured that 90\% of the light collected is transmitted through the coloured glass and bandpass filter setup. The spatial filter results in a 30\,dB reduction in the background scattered light with minimal reduction in the atomic fluorescence signal. The solid angle of the collected light is $\approx 4\times 10^{-2}$\,sr. The PMT signal is fed into a lock-in amplifier and the output of the lock-in amplifier is digitized with an oscilloscope (Tektronix TBS2104) along with the signal from the reference photodiode. The reference signal used for the lock-in comes from an optical chopper (ThorLabs) used for the probe beam. The fluorescence signal is normalized by the signal from the reference photodiode to reduce the effect of laser-power variations.

\begin{figure}[t]
    \centering
    \includegraphics[width=\columnwidth]{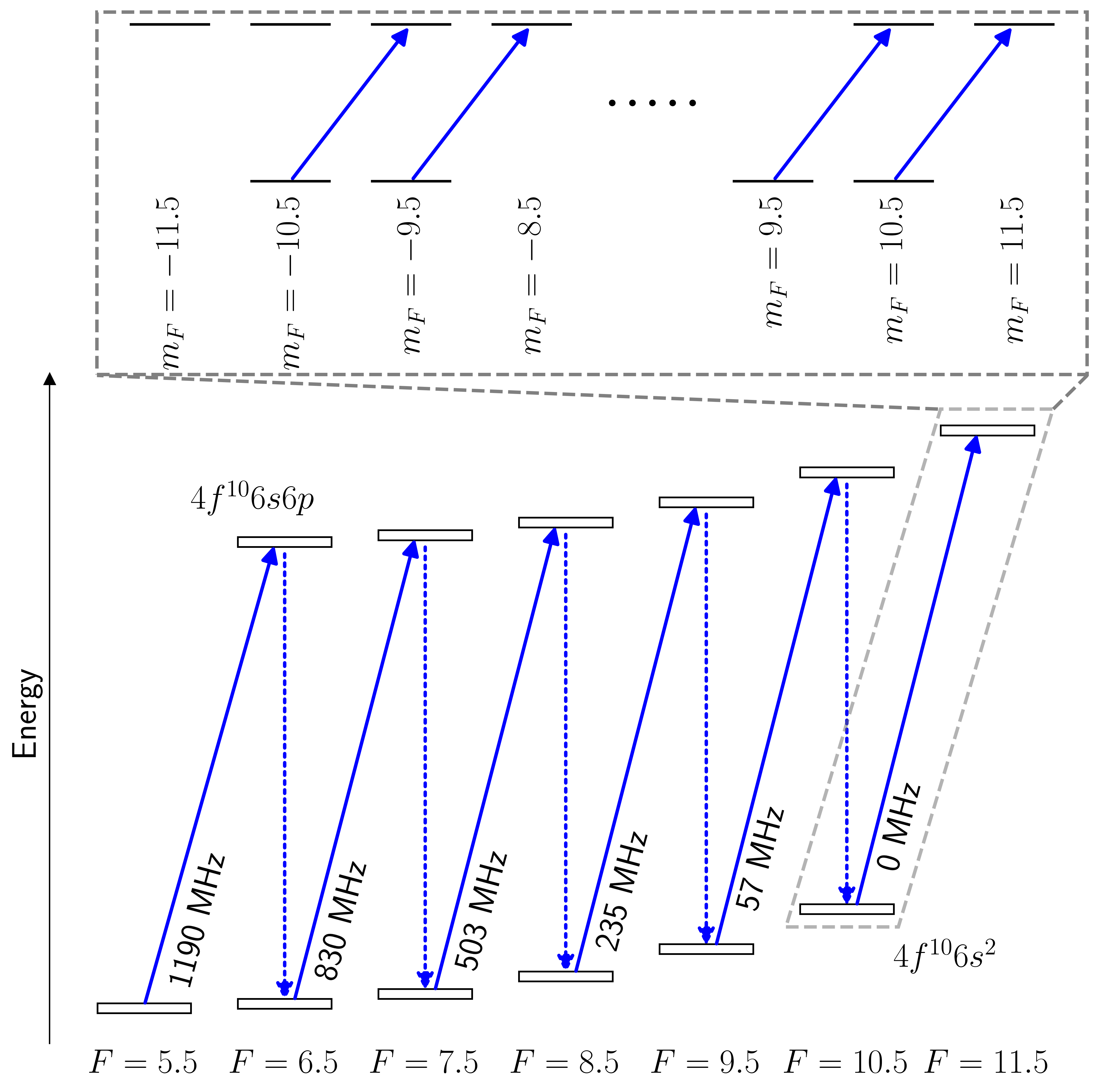}

    \caption{\label{fig:hyperfine structure} Schematic for optical pumping of $^{163}$Dy on the 421\,nm $J=8\rightarrow J=9$ transition. Once excited to an $F$ hyperfine state of the excited state, the atoms can generally decay to the $F-1, F$ and $F+1$ ground hyperfine states (if these are available), with the latter being the weakest due to the $\Delta F = \Delta J$ rule \cite{auzinsh2010opticallypolarisedatoms}. We only show the $\Delta F=0$ decays in the figure.}
\end{figure}

\subsection{Electro-Optic Modulator}

A custom broadband electro-optic modulator (QUBIG) is used to generate the frequency sidebands required for the hyperfine pumping of the ${}^{163}$Dy isotope following an idea previously proposed in\,\cite{Leefer2010_Transverse_cooling}. The EOM uses a lithium niobate crystal with an anti-reflection coating for 421\,nm, the insertion losses typically range between 0.5 and 0.7\,dB. The half-wave voltage of the EOM is $V_\pi=76$\,V. The RF drivers and mixer were designed and provided along with the EOM by QUBIG. The control unit consists of five RF oscillators tuned to the relevant frequencies, each with their own amplifier. They are then sent to the EOM via a mixer. The pump laser is locked to the ${}^{163}$Dy $F=10.5 \rightarrow F'=11.5$ hyperfine component of the 421\,nm transition and five sidebands with frequency detunings of 1190\,MHz, 830\,MHz, 503\,MHz, 235\,MHz and 57\,MHz are generated. The schematic for optical pumping of these hyperfine and Zeeman states in $^{163}$Dy is shown in Fig.\,\ref{fig:hyperfine structure}. The spectrum generated with the EOM is analyzed using a Fabry-Perot spectrum analyzer with a free spectral range of $\simeq 3900$\,MHz and finesse of $\simeq 700$.

\begin{figure}[b]
    \includegraphics[width = \columnwidth]{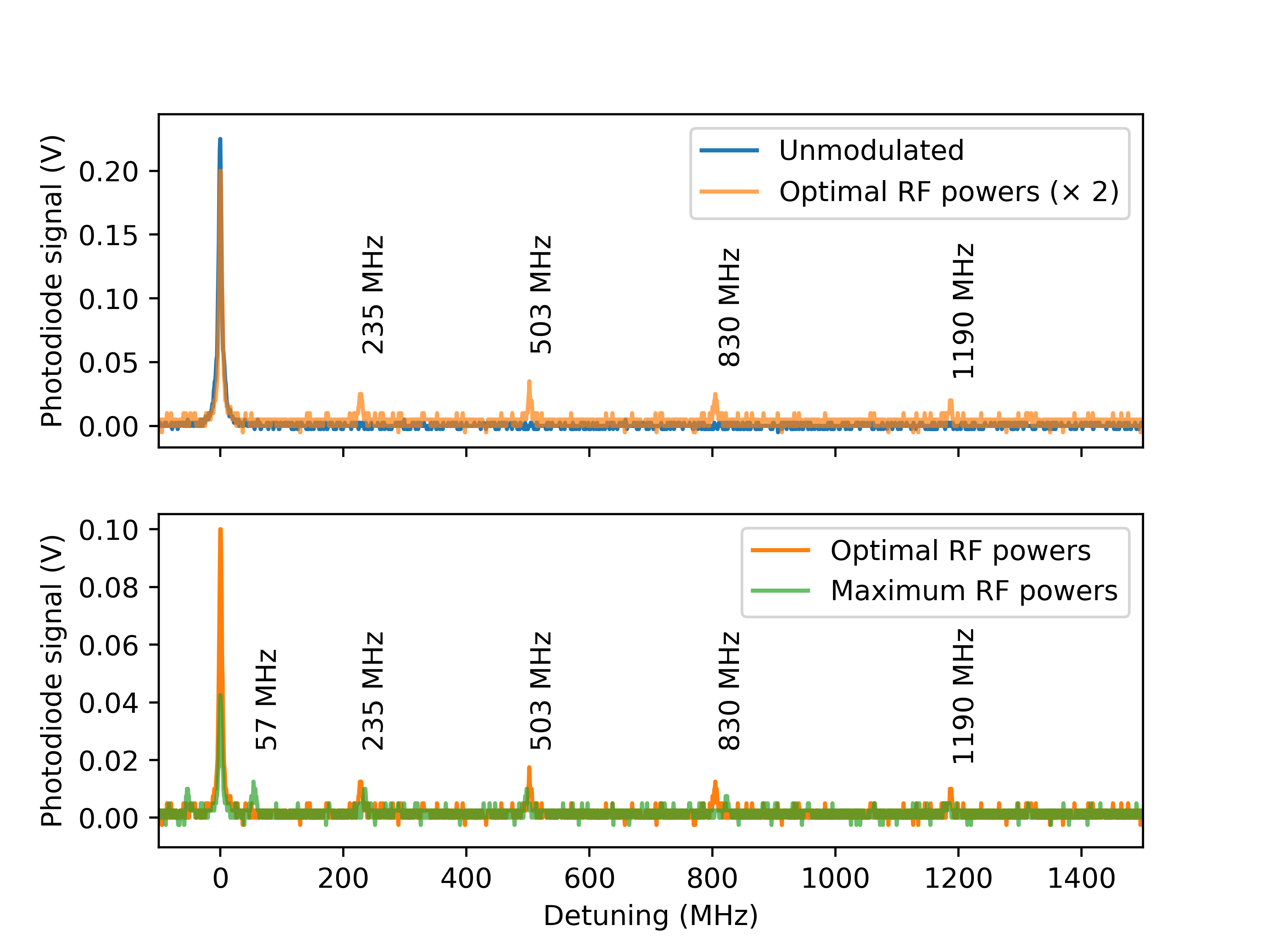}

    \caption{\label{fig:EOMspectra} Fabry-Perot spectrum analyzer signal with no modulation, optimal-power modulation for optical pumping and maximum RF powers for modulation (see text).
    }
\end{figure}

\begin{table}
\begin{ruledtabular}
\begin{tabular}{ccc}
\rule{0pt}{2.5ex} Detuning & Optimal RF Power& Hyperfine Component\\
\rule{0pt}{2.5ex} (MHz) &(dBm)  & ($J = 8 \rightarrow J' = 9$)\\[1ex]
\hline 
\rule{0pt}{2.5ex} 1190 & 32 & $F = 5.5 \rightarrow F' = 6.5$\\
\rule{0pt}{2.5ex} 830 & 32 & $F = 6.5 \rightarrow F' = 7.5$\\
\rule{0pt}{2.5ex} 503 & 33 & $F = 7.5 \rightarrow F' =  8.5$\\
\rule{0pt}{2.5ex} 235 & 31 & $F = 8.5 \rightarrow F' = 9.5$\\
\rule{0pt}{2.5ex} 57 & 15.5 & $F = 9.5 \rightarrow  F' = 10.5$\\
\rule{0pt}{2.5ex} 0 & - & $F = 10.5 \rightarrow  F' = 11.5$\\
\end{tabular}
\end{ruledtabular}
\caption{\label{tab:RFPower} RF powers used to generate the sidebands required for efficient hyperfine pumping in {$^{163}$Dy}. Zero detuning corresponds to the frequency of the $F=10.5 \rightarrow F'=11.5$ transition. }
\end{table}

Fig.\,\ref{fig:EOMspectra} shows the spectra obtained with the Fabry-Perot spectrum analyzer and the frequency detunings corresponding to all five frequencies being observed. The RF powers used to generate the sidebands are listed in Table\,\ref{tab:RFPower}. A decrease in the laser power in the carrier and sidebands is observed when the EOM is driven at the maximum allowed RF powers of 34\,dBm for all modulation frequencies. The 57\,MHz modulation is not driven at high RF powers for observation of optimal hyperfine optical pumping due to the strong off-resonant coupling of the pump laser to the $F=9.5 \rightarrow F'=10.5$ hyperfine transition when it is locked to the $F=10.5 \rightarrow F'=11.5$ transition. This off-resonant coupling is observed because of the small frequency shift of 57\,MHz ($< 2\Gamma$) between the two transitions due to hyperfine interactions.

\section{Experimental results}

The frequency of the 421\,nm pump laser is tuned to a relevant Dy transition for optical pumping and is appropriately detuned to observe optimal simultaneous laser cooling and optical pumping. The fluorescence signal from the atoms along with the reference photodiode signal are recorded as the probe laser is scanned over the isotope and hyperfine components of the 599\,nm transition (scan range of 6\,GHz; duration of scan 5\,s); and example of a scan can be found in Sec.\ref{Hyperfineandzeeman}, Fig.\,\ref{fig:161 ZeemanandHyperfine}.
Ten different scans are recorded for each spectrum for averaging. The prominent spectral lines are those of $^{164}$Dy (natural abundance 28.18\,\%), $^{163}$Dy (24.9\,\%), $^{162}$Dy (25.5\,\%) and $^{161}$Dy (18.9\,\%). The $^{163}$Dy and the $^{161}$Dy isotopes both have nuclear spin 5/2 so their spectral lines are split by hyperfine interactions. A bias magnetic field $\simeq$ 1.5\,G is applied along the direction of propagation of the pump laser beam (chosen as the quantization axis) using the coils inside the optical pumping region.

The reference photodiode signal is used to normalize the fluorescence signal. The position of the $^{164}$Dy peak is set as the zero of the $x$-axis and the detuning is calibrated using the isotope shift between the Dy isotopes found in Ref.\,\cite{Childs1977_598nm}. The other prominent peaks are identified using the isotope shifts and the hyperfine structure coefficients of the 599\,nm transition found in Ref.\,\cite{Childs1977_598nm}.

The linewidth of the $^{163}$Dy $F = 10.5 \rightarrow F'=9.5$ transition is obtained from the fit and the FWHM linewidth is observed to be $\simeq$ 32\,MHz. The natural linewidth of the transition is 12\,kHz\,\cite{NIST_ASD}. The larger observed width is due to the residual Doppler broadening arising from atomic beam divergence which leads to characterized transverse velocities of $13$\,m/s. 

To achieve our goal of efficient simultaneous Zeeman and hyperfine pumping as well as laser cooling of an atomic beam of Dy, we first need to prepare the internal state of the Dy atoms employing the 421\,nm laser and the multitone EOM as discussed above and then assess this preparation to optimize it via our experimental parameters. Those parameters include the frequency and polarization state of the 421\,nm light and the relative optical power in the sidebands.

\subsection{Polarization Optimization}

\begin{figure}[t]
    \includegraphics[width = \columnwidth]{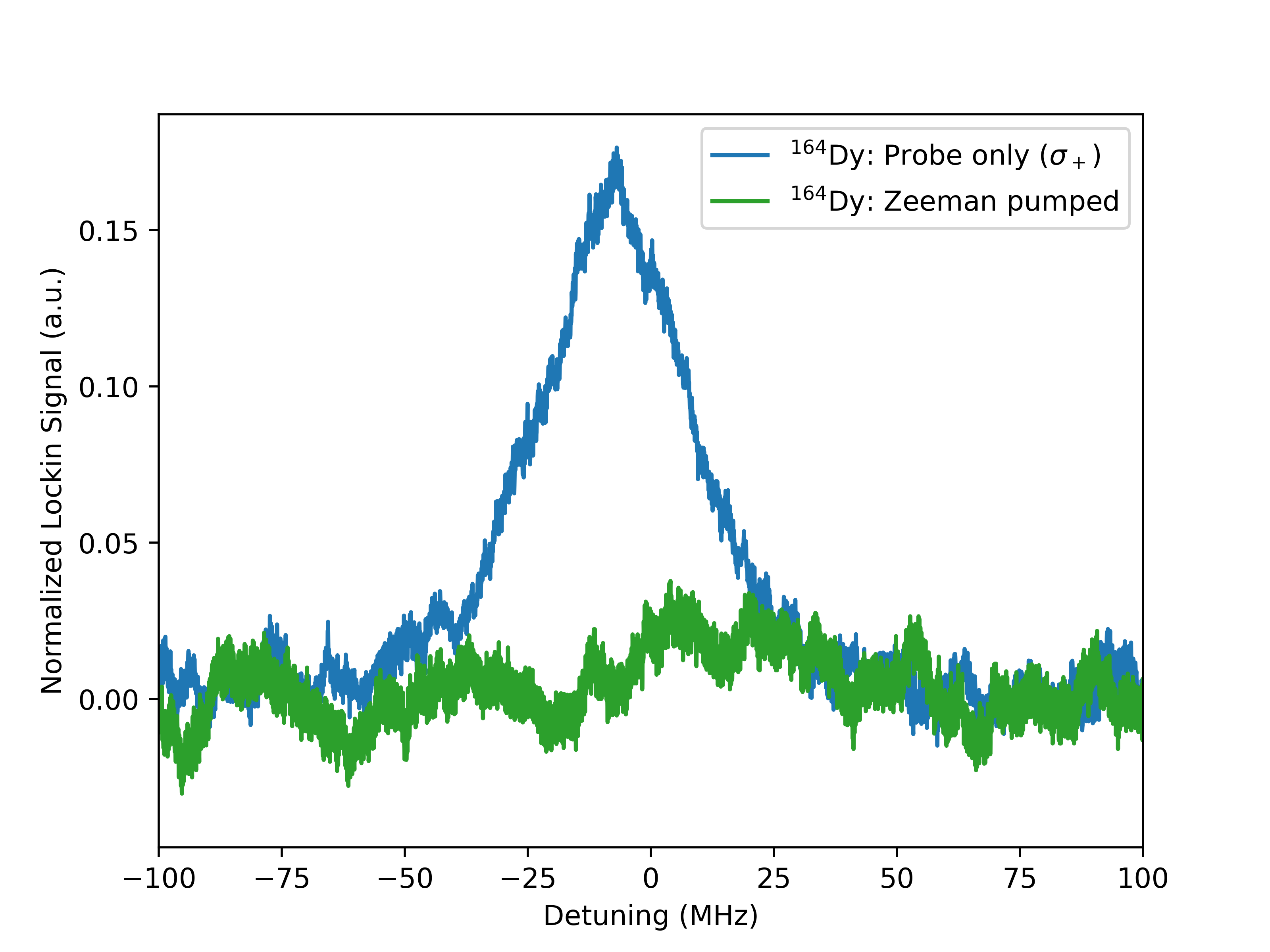}
    \caption{\label{fig:Polarization optimization} Fluorescence $\sigma_+$ probe spectra with pump powers of 0\,mW and 230\,mW. A vanishing $^{164}$Dy spectral line indicates optimized polarization of the pump laser. Residual oscillations of amplitude in the probe spectrum are observed after optical pumping.}
\end{figure}

To optimize the polarization of the 421\,nm pump laser, the probe polarization was tuned to induce $\sigma_+$ transitions and the polarization of the pump laser is controlled to induce $\sigma_+$ transitions. The pump laser frequency $\omega_o$ is tuned to the $^{164}$Dy $J=8\rightarrow J'=9$ transition. Since the 599\,nm probe is a $J=8 \rightarrow J'=7$ transition, an atomic beam optimally pumped to the $J=8,$ $m_J=8$ ground state in $^{164}$Dy would be a dark state for a $\sigma_+$ probe. A vanishing $^{164}$Dy spectral line in the probe would thus indicate that the pump polarization is successfully set to transfer the ground state population fully into the $m_J=8$ sublevel. As seen in Fig.\,\ref{fig:Polarization optimization} this is experimentally achieved by careful alignment of the quarter wave plate. This shows optimal Zeeman pumping of the atomic beam but since the atoms are pumped to a dark state of the probe, it is impossible to discern the degree of laser cooling. Therefore we turn the probe quarter wave plate 90$^\circ$ to induce $\sigma_-$ transition and interrogate the $m_J=8$ population which allows us to measure the degree of laser cooling and optical pumping of the atomic beam.

\subsection{Zeeman Optical Pumping and Laser Cooling}

To observe simultaneous laser cooling and Zeeman optical pumping, the pump laser is first tuned to the $^{164}$Dy $J=8\rightarrow J'=9$ transition. An arbitrary detuning of $0<\delta<\Gamma/2$ is initially chosen and is experimentally tuned to observe optimal optical pumping. The increase in the population of the atoms in the $J=8$, $m_J=8$ Zeeman sublevel as well as the degree of laser cooling are detected by observing the increase in the amplitude of the spectral line corresponding to the $^{164}$Dy $J=8\rightarrow J'=7$ probe spectrum. 

\begin{figure}[t]
    \includegraphics[width = \columnwidth]{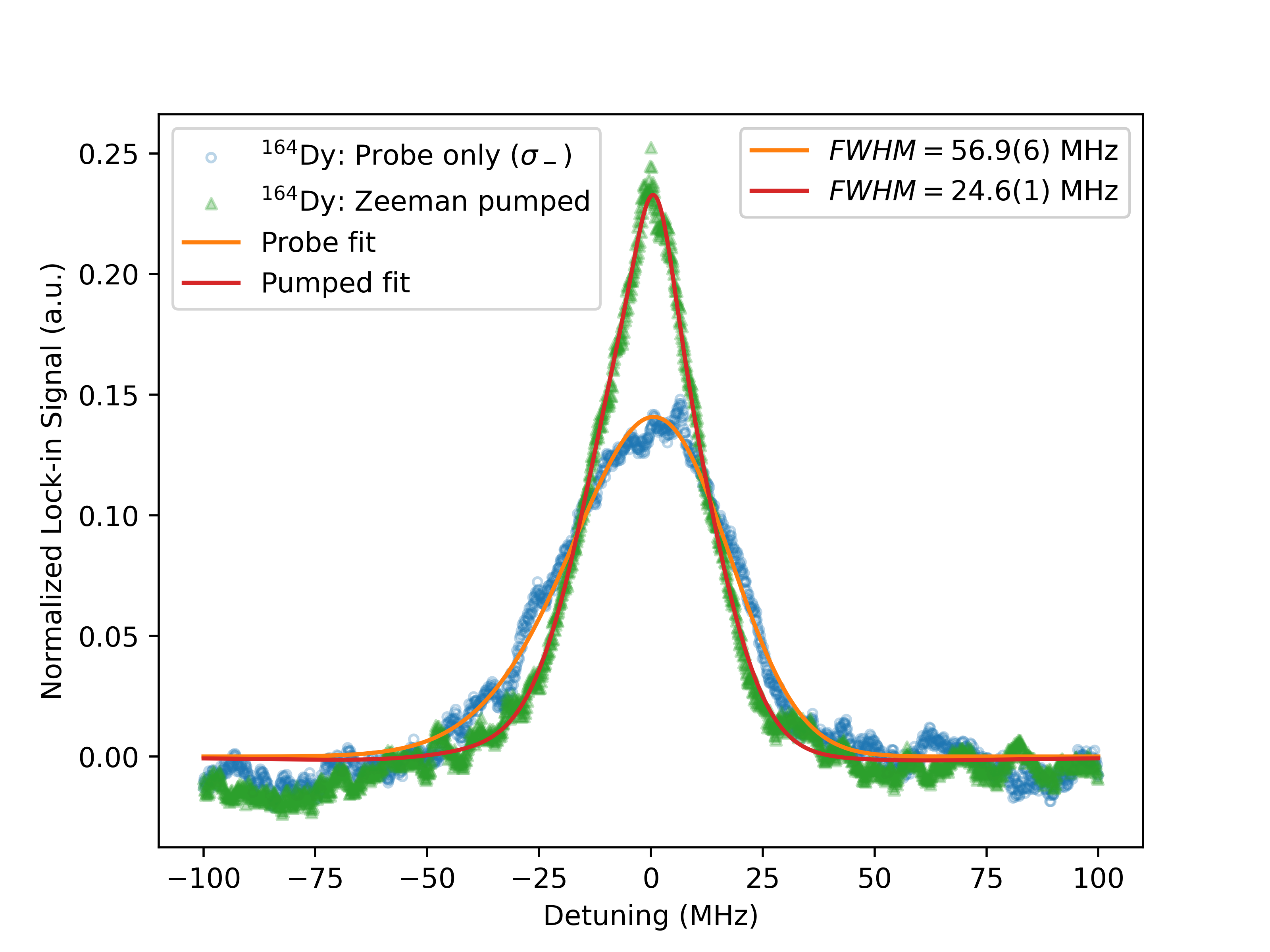}
    \caption{\label{fig:164 Zeeman} Fluorescence $\sigma_-$ probe spectra before and after Zeeman optical pumping with a $\sigma_+$ pump laser tuned to $^{164}$Dy $J=8 \rightarrow J'=9$ transition with a power of 230\,mW.}
\end{figure}

Multiple spectra obtained are interpolated to a common detuning and averaged. A nonlinear least squares fit is performed on the normalized and averaged fluorescence spectra of the well resolved $\ket{F=10.5}\rightarrow\ket{F=9.5}$ transition. The probe spectrum without the pump laser is fit using a Gaussian profile and the optically pumped spectra are fit with a sum of Gaussian and Voigt profiles. The Gaussian and Voigt profiles are of the form
\begin{equation}\label{Gauss}
    f_{G}(f) = \frac{A}{\sigma_g \sqrt{2\pi}}e^{-(f-f_o)^2/2\sigma_g^2},    
\end{equation}
\begin{equation}\label{Voigt}
    f_{V}(f) = \int^\infty_{-\infty} f_{G}(f) \frac{1}{\pi} \frac{\gamma_v/2}{(f-f_o)^2 + \gamma_v^2/4}df_o,
\end{equation}
where $\sigma_g$ and $\gamma_v$ are the Doppler width of the spectrum and width at FWHM due to the laser frequency spectrum, respectively. From Eq.\,\ref{Voigt}, $\sigma_g$ is used to determine the degree of laser cooling. An example of these fits can be found in Fig.\,\ref{fig:164 Zeeman} which shows the probe spectrum before and after Zeeman pumping and cooling. The $^{164}$Dy probe spectrum is amplified by a factor of $1.7(2)$ which is greater than the theoretical estimate (Appendix\,\ref{Estimation}) due to laser cooling of the atomic beam. The uncertainty shown here and in the rest of the letter is the uncertainty in the fits. The FWHM width of the pumped and cooled spectrum obtained from the fit is $24.6(1)$\,MHz. The FWHM of the uncooled probe spectrum is $56.9(6)$\,MHz.

\begin{figure}[b]
    \includegraphics[width = \columnwidth]{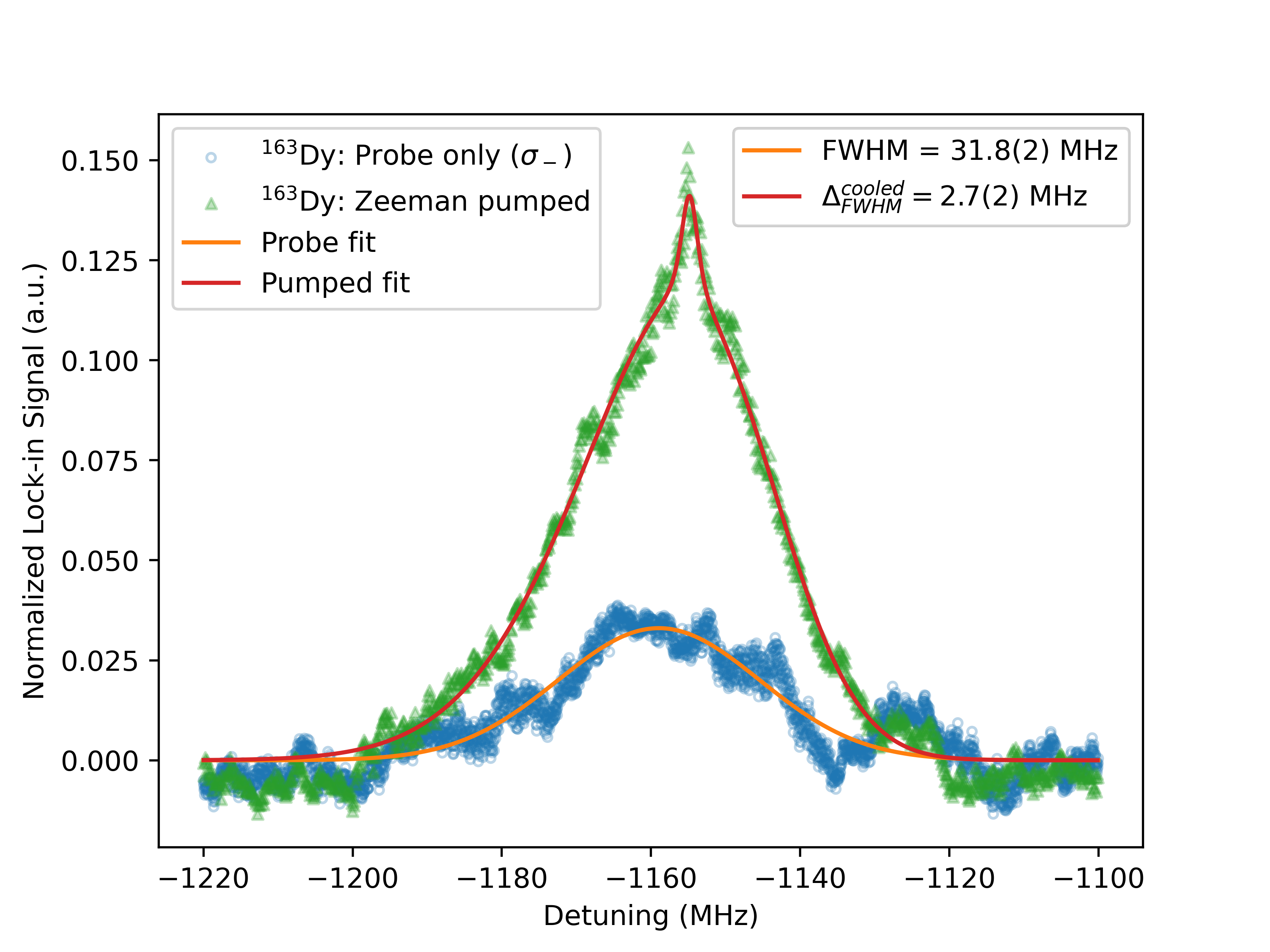}
    \caption{\label{fig:163 Zeeman} Fluorescence $\sigma_-$ probe spectra with and without Zeeman optical pumping and cooling with a $\sigma_+$ pump laser tuned to $^{163}$Dy $F=10.5 \rightarrow F'=11.5$ hyperfine transition with a power of 230\,mW.}
\end{figure}

The pump laser is then tuned to the $^{163}$Dy $F=10.5\rightarrow F'=11.5$ hyperfine transition and the increase in the amplitude of the $F=10.5 \rightarrow F'=9.5$ spectral line in the probe spectrum is observed. Fig. \ref{fig:163 Zeeman} shows the probe spectrum with and without Zeeman pumping and laser cooling the $^{163}$Dy isotope. The amplification in the amplitude by a factor of $4.2(1)$ is observed in the probe spectrum. Assuming Zeeman pumping of the $F = 10.5$ hyperfine ground state only, we expect an amplification factor of 1.2. The observed gain is higher than the expected one because of off-resonant hyperfine pumping from the $F = 9.5$ hyperfine ground state that brings more population into the $F = 10.5$ hyperfine ground state as well as laser cooling the atomic beam, both of which lead to a larger fluorescence signal. The off-resonant optical pumping from the $F=9.5$ hyperfine ground occurs due the $\ket{F=9.5} \rightarrow \ket{F=10.5}$ transition being less than $2 \Gamma$ away from the $\ket{F=10.5} \rightarrow \ket{F=11.5}$ transition.

The expected Doppler width of the $J=8 \rightarrow J=7$ probe transition at the Doppler cooling limit due to the 421\,nm pump is $\sigma_d = \nu_o/c\sqrt{\hbar/2M\tau}\sqrt{\ln{2}}=0.28$\,MHz, where $\nu_o$ is the frequency of the transition, $M$ the mass of dysprosium, and $\tau$ the lifetime of the state. The Doppler FWHM of the cooled profile of the $^{163}$Dy spectrum obtained from the fit to Eq.\,\ref{Voigt} is $2.7(2)$\,MHz and $\sigma_g = 0.59(39)$\,MHz and $\gamma_v=0.99(43)$\,MHz. The obtained $\sigma_g \simeq 0.6$\,MHz values are of the same order as the expected broadening at the Doppler limit. Any residual broadening in the probe spectrum is due to the laser linewidth of 1\,MHz.

\subsection{Hyperfine and Zeeman Optical Pumping and Laser Cooling}\label{Hyperfineandzeeman}

\begin{figure}[t]
    \includegraphics[width = \columnwidth]{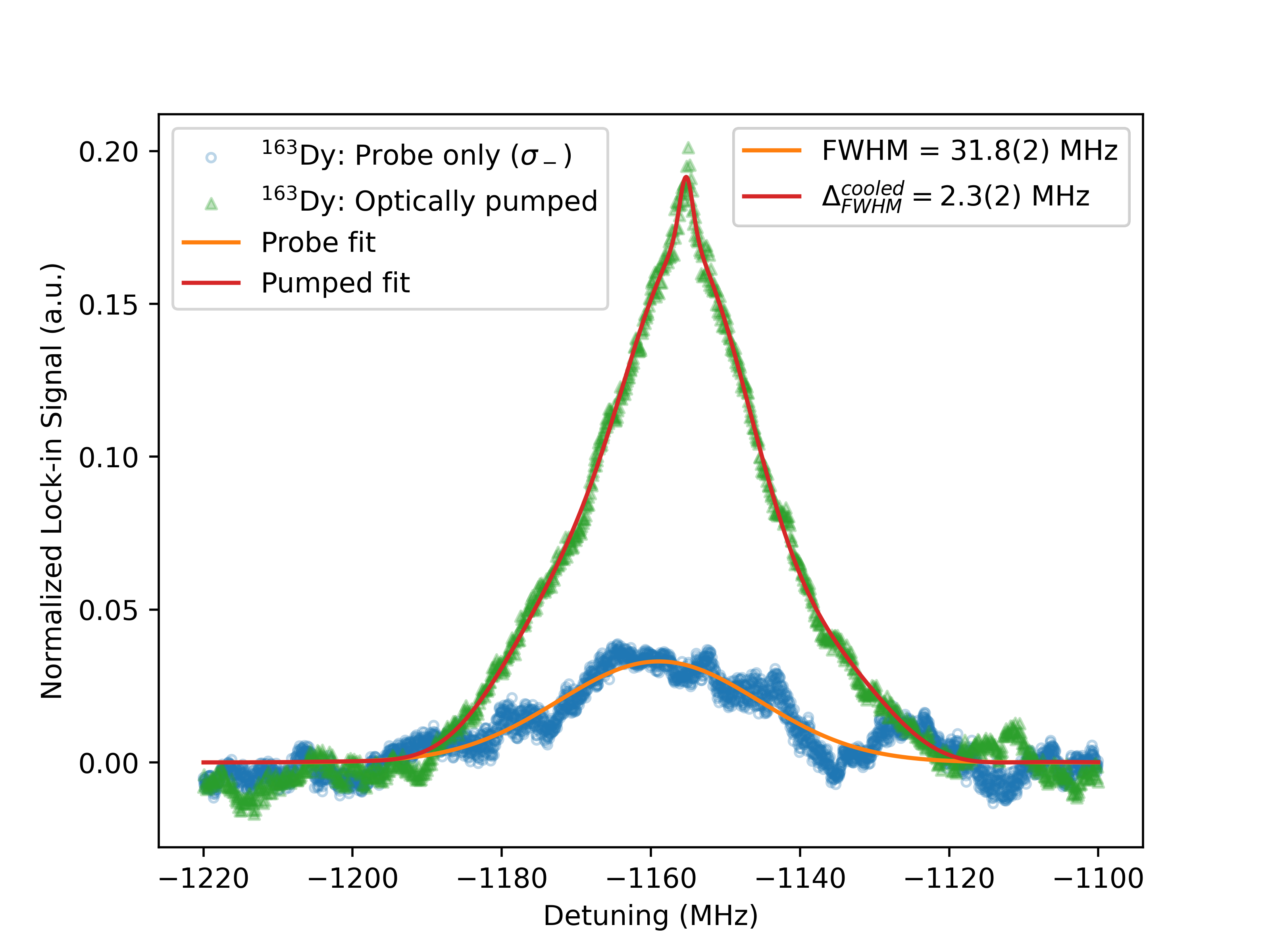}
    \caption{\label{fig:163 ZeemanandHyperfine} Fluorescence $\sigma_-$ probe spectra with and without hyperfine and Zeeman optical pumping and cooling with a $\sigma_+$ pump laser tuned to $^{163}$Dy $F=10.5 \rightarrow F'=11.5$ hyperfine transition with a power of 230\,mW and EOM RF modulation powers shown in Table\,\ref{tab:RFPower}.}
\end{figure}

\begin{figure}[b]
    \includegraphics[width = \columnwidth]{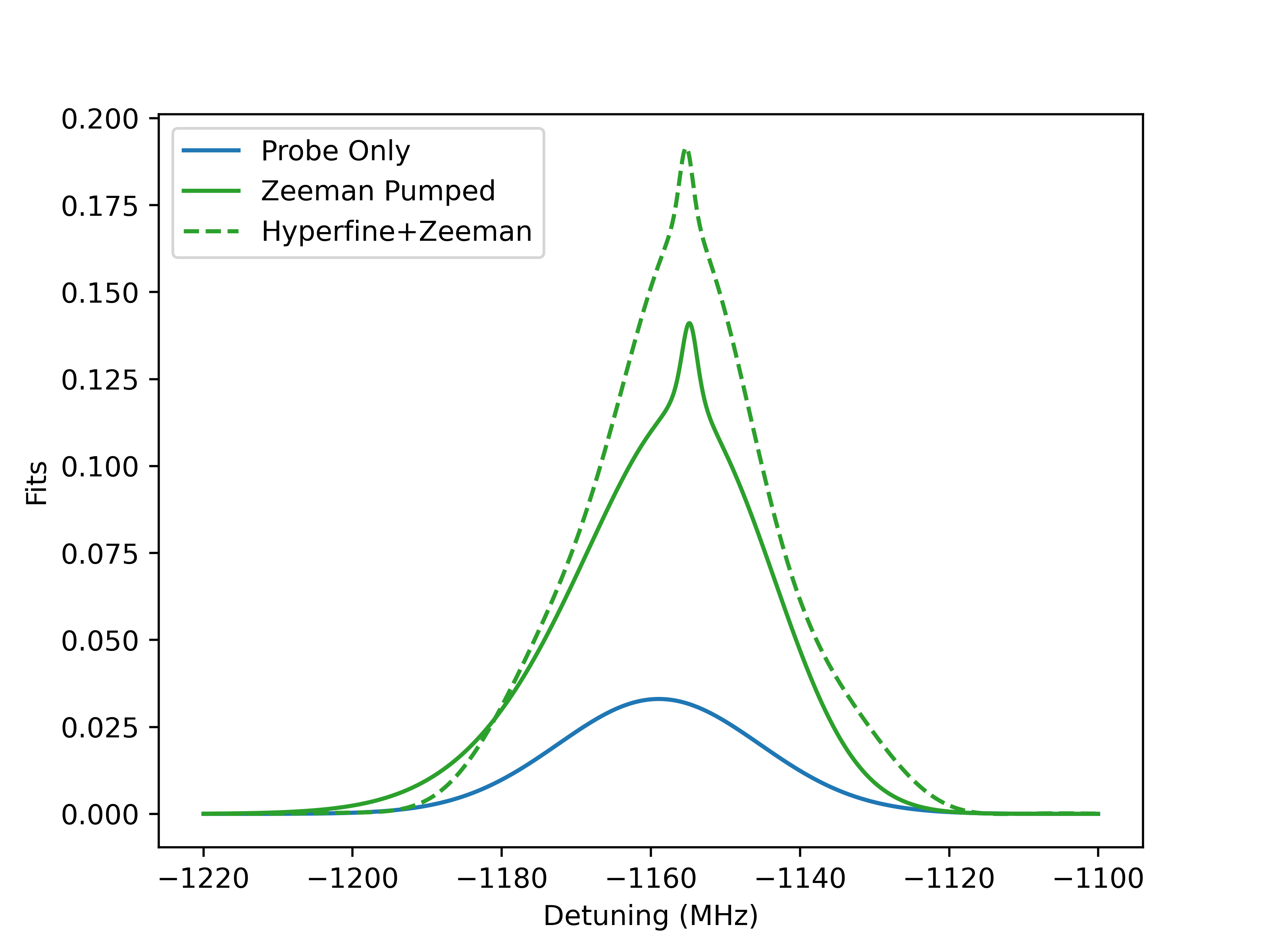}
    \caption{\label{fig:FitComparison} Comparison of the fits of the Zeeman pumped, hyperfine and Zeeman pumped and probe spectra.}
\end{figure}

\begin{figure}[t]
    \includegraphics[width = \columnwidth]{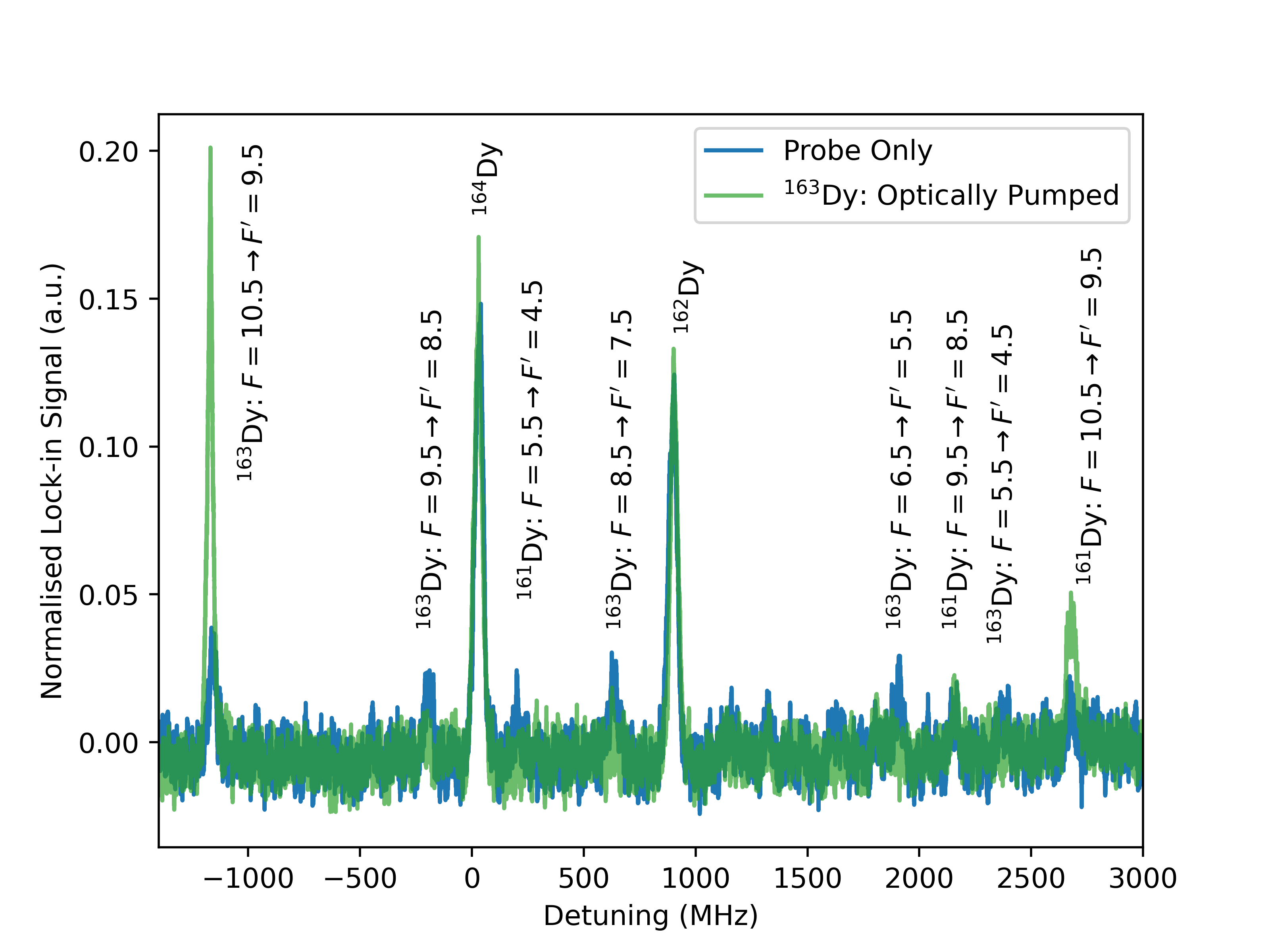}
    \caption{\label{fig:161 ZeemanandHyperfine} Accidental hyperfine and Zeeman optical pumping and cooling of the $^{161}$Dy isotope with the $\sigma_+$ pump laser tuned to $^{163}$Dy $F=10.5 \rightarrow F'=11.5$ hyperfine transition with a power of 230\,mW and EOM RF modulation powers shown in Table\,\ref{tab:RFPower}.}
\end{figure}

To investigate hyperfine and Zeeman optical pumping and laser cooling of the $^{163}$Dy isotope, the pump laser is locked to the $F=10.5 \rightarrow F'=11.5$ hyperfine transition and the EOM is used to modulate the pump laser frequency. The pump laser frequency is then red-shifted to observe optimal laser cooling and optical pumping. The pump laser polarization is tuned to induce $\sigma_+$ transitions and the probe is tuned to induce $\sigma_-$ transitions. An increase in the amplitude of the $F=10.5 \rightarrow F'=9.5$ hyperfine transition of the probe is observed. 

Fig.\,\ref{fig:163 ZeemanandHyperfine} shows the probe spectrum with and without optically pumping the $^{163}$Dy isotope and a comparison of the fits of the Zeeman pumped and hyperfine and Zeeman pumped $^{163}$Dy $F=10.5 \rightarrow F'=9.5$ spectrum is shown in Fig.\,\ref{fig:FitComparison}. An amplification of $5.9(7)$ is observed in the  $F=10.5 \rightarrow F'=9.5$ hyperfine transition of the probe with hyperfine and Zeeman optical pumping. This is in agreement with the theoretical expectation of a factor of 5.4 which indicates that almost all the atoms in the atomic beam have been pumped to the $\ket{F=10.5, m_F=10.5}$ ground state. The optical pumping of atoms in the other hyperfine ground states is also made evident by the lack of spectral lines corresponding to the transitions of the optically pumped hyperfine ground states in Fig.\,\ref{fig:161 ZeemanandHyperfine}.

The FWHM of the cooled profile of $^{163}$Dy spectrum obtained from the fit to Eq.\,\ref{Voigt} is $2.3(2)$\,MHz with $\sigma_g = 0.61(39)$\,MHz and $\gamma_v = 0.7(5)$\,MHz. This is roughly in agreement with the Doppler cooling limit. The $\gamma_v$ values obtained in both the Zeeman pumped and hyperfine and Zeeman pumped spectra agree. Not all the atoms here are cooled to the Doppler limit since the atoms getting hyperfine pumped to the $F=10.5,$ $m_F=10.5$ state do not cycle in the $F=10.5$, $m_F=10.5 \rightarrow F=11.5$, $m_F=11.5$ cooling transition long enough to be cooled to the Doppler limit. This requires the data to be fit with a sum of a Gaussian profile and two Voigt profiles.

Accidental hyperfine and Zeeman optical pumping of the $^{161}$Dy isotope was also observed with the pump laser tuned to the $^{163}$Dy $F=10.5 \rightarrow F'=11.5$ hyperfine transition. This was due to resonances of the frequency sidebands generated by the EOM with the $^{161}$Dy hyperfine transitions. Fig.\,\ref{fig:161 ZeemanandHyperfine} shows the amplification in the $^{161}$Dy $F=10.5 \rightarrow F'=9.5$ hyperfine transition in the probe spectrum. An amplification factor of $3.0(7)$ was observed in the $^{161}$Dy $F=10.5 \rightarrow F'=9.5$ probe spectrum with no laser cooling observed. This is because the closest frequency sideband generated by the EOM is detuned by $\simeq 700$\,MHz from the $F=10.5 \rightarrow F'=11.5$ hyperfine component of the $^{161}$Dy transition.


\section{Conclusions and Outlook}

Efficient simultaneous optical pumping and laser cooling of atomic dysprosium was achieved with the use of an EOM and a single laser beam. The amplification of the amplitude of the $F=10.5 \rightarrow F'=9.5$ peak of ${}^{163}$Dy in the probe signal after optical pumping is a clear indication that atoms in the hyperfine states were optically pumped into the $F = 10.5$ hyperfine state and the reduced width of the probe spectrum is a clear indicator of laser cooling. Furthermore, a comparison of experimentally obtained amplification in the amplitude of probe spectra with theoretical estimates indicates optimal hyperfine and Zeeman optical pumping. The experimentally obtained laser linewidths indicate that the atomic beam has been cooled to the Doppler limit. This work has been performed on a atomic beam of Dy but the present technique of using a multitone EOM modulation to pump the hyperfine states of Dy can in principle be used in other configurations such as trapped atoms, molecules and ions where one could expect 100\% efficiency in optically pumping the atoms to the polarized states.

The optically pumped dysprosium atoms will be used in the search for parity violation in dysprosium. The previous search for parity violation (PV) \cite{Nguyen1997_PNC_Dy}, the most sensitive parity-violation search so far in terms of the weak-mixing amplitude, has reached into the expected range of the effect but yielded a result consistent with zero. The boost in sensitivity, which is proportional to the number of atoms in the $\ket{F=10.5, m_F=10.5}$ state, by a factor of $102$ afforded by the results of the present work may allow the detection of the long-sought PV signals in Dy.


\begin{acknowledgments}

This research was supported in part by the German Research Foundation (DFG), project ID 390831469: EXC 2118 (PRISMA+ Cluster of Excellence) and project ID BU3035/10-1. JBK is grateful for the support of the National Research Foundation of Korea (NRF-2021R1F1A1060385). JA is grateful for the support of the Humboldt Foundation. D. Antypas contributed to the project in its early stages. RC is grateful for the continued support and guidance of DB.

\end{acknowledgments}


\appendix

\section{Theory and Simulation}\label{AppA}

$^{163}$Dy has a nuclear spin of 5/2 which leads to a hyperfine structure with six hyperfine states with quantum numbers $F=5.5$ to $10.5$ in the ground $J=8$ state. Each hyperfine state is further split into $2F+1$ Zeeman sublevels that are degenerate in the absence of an external magnetic field. Pumping the atoms in these hyperfine and Zeeman sublevels to the $F=10.5,\,m_F = 10.5$ Zeeman sublevel requires pumping the atoms in the $F=5.5$ to $9.5$ ground hyperfine states to the $F=10.5$ ground hyperfine state while controlling the polarization of the pump laser to induce $\sigma_+$ transitions.

The optical pumping dynamics can be described by the Liouville-von Neumann equations\,\cite{auzinsh2010opticallypolarisedatoms}
\begin{equation}
\frac{d\rho}{dt} = -\frac{i}{\hbar}[H, \rho] - \frac{1}{2}\{\Gamma, \rho\} + \Lambda\,,    
\end{equation}
where $\rho$ is the density matrix of the system and $H$ is its Hamiltonian, $\Gamma$ is a diagonal matrix describing the spontaneous decay from the excited state, and $\Lambda$ is the matrix describing the rate of repopulation of the ground state because of spontaneous decay from the excited state. The broad linewidth of the 421\,nm transition leads to off-resonant coupling and hyperfine optical pumping from the $F=9.5$ ground hyperfine state to the $F=10.5$ ground state while the pump laser is locked to the $F=10.5 \rightarrow F'=11.5$ transition making it a poor candidate to test the extent of Zeeman optical pumping. Thus, $^{164}$Dy is chosen to optimize the polarization of pump laser light for optimal Zeeman pumping due to the lack of nuclear spin and thus a lack of hyperfine structure.

\begin{figure}[t]
    \centering
    \includegraphics[width=\columnwidth]{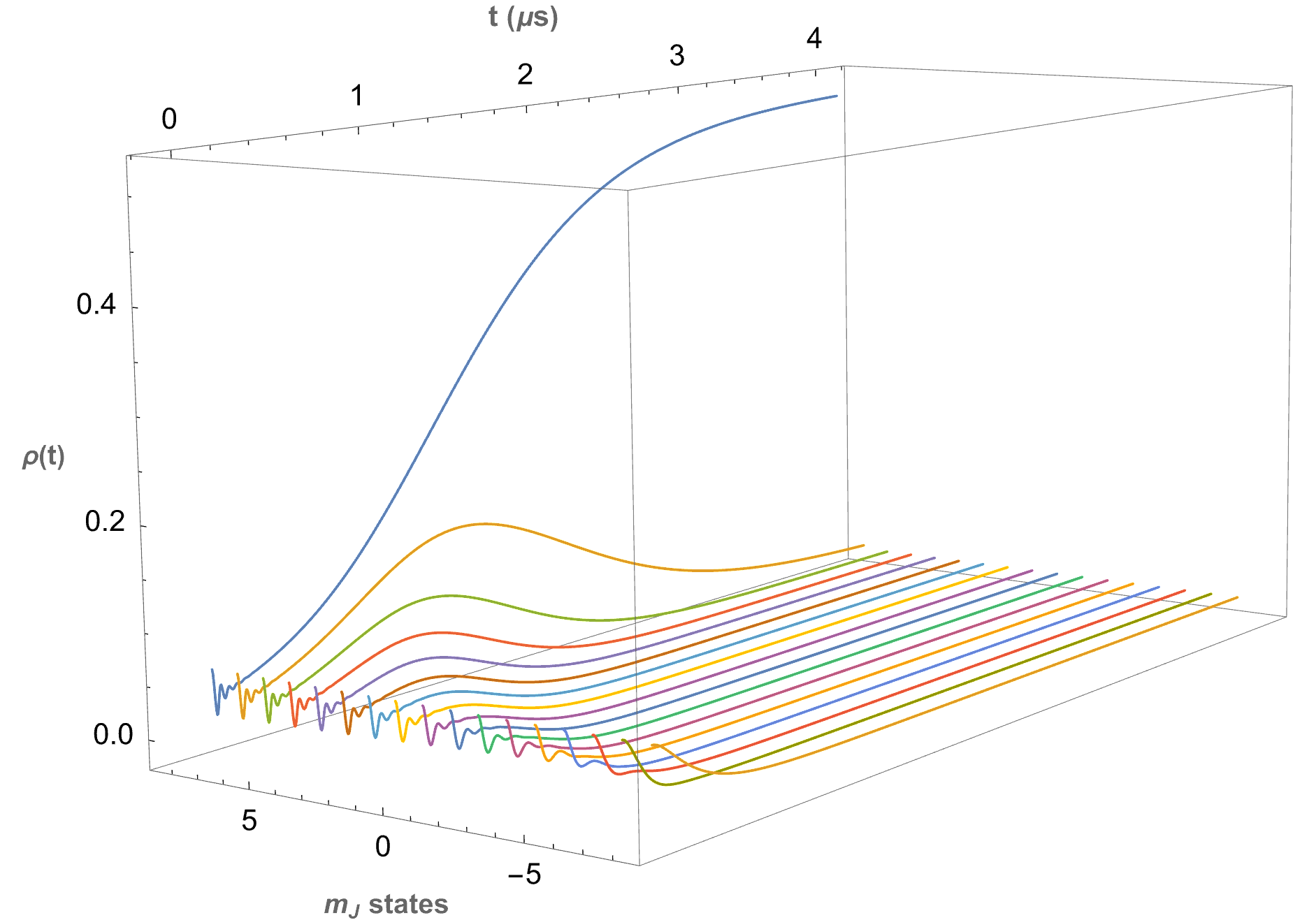}
    \caption{\label{fig:Zeeman simulation} Time evolution of the populations in the Zeeman sublevels of $^{164}$Dy isotope with $\sigma_+$ transitions driven by the pump laser. Simulations assume a total population, $\sum_{m=-J}^JP(m) = 1$, distributed equally among all Zeeman sublevels. The steady state shows a population of 0.5 in the $m_J = J$ ground state since the other half of the population is in the excited state, $m_{J'} = J'$.}
\end{figure}

Simulations\,\footnote{Simulations are performed using the AtomicDensityMatrix package on Mathematica} of Zeeman optical pumping of $^{164}$Dy are performed to estimate the time taken to reach a steady state and completely pump the system to the $m_J = J$ state. Simulations are performed with laser powers used in the experiment and show that the atoms are Zeeman pumped to the $m_J=8$ state in $\simeq 4$\,$\mu$s. The simulations assume no Doppler broadening and degeneracy of the Zeeman sublevels. Fig.\,\ref{fig:Zeeman simulation} shows the time evolution of the population of the ground state Zeeman sublevels at a laser power of 230\,mW. Since the experimental interaction times are of the order of 65\,$\mu$s, it is safe to assume that we are fully Zeeman pumping the atomic beam.

\section{Estimation of Degree of Optical Pumping}\label{Estimation}

The experimental scheme for measuring the degree of optical pumping relies on measuring the change in the signal strength of the spectral line associated with the optically pumped atoms. The signal strength due to laser-induced fluorescence is proportional to the population in the excited state and the transition probability of the excited state to the ground state. 

The excited state population is given by\,\cite{van_der_Straten_Metcalf_2016},
\begin{equation}\label{excited state population}
    \rho_{ee} = \frac{s/2}{1+s+(2\Delta/\Gamma)^2}\,,
\end{equation}
where $\Delta$ is the detuning of the laser from the atomic transition frequency and $\Gamma$ is the lifetime of the excited state and $s$ is the saturation parameter for the transition on resonance. The saturation parameter for a $\ket{F, m_F} \rightarrow \ket{F', m_{F'}}$ transition is given by\,\cite{van_der_Straten_Metcalf_2016},
\begin{equation}\label{saturation}
    s = \Omega^2/\Gamma^2 = \frac{\braket{F', m_{F'}|\textbf{D}\cdot \textbf{E}|F, m_F}^2}{\Gamma^2},
\end{equation}
where $\Omega^2$ is the rate of excitation, $\Gamma$ is the rate of decay from the excited state. The observed fluorescence due to the decay of an excited $m_F$ state would then be proportional to the sum of the transition probabilities through all possible decay paths, and due to the experimental geometry the fluorescence of the $\Delta m_F=q=0$ decay is maximally observed using the PMT. The observed fluorescence is then given by,
\begin{equation}
    \mathcal{F} \simeq \rho_{ee} \times \sum_{m_F} \braket{F, m_F|\textbf{D}\cdot \textbf{E}|F', m_{F'}}^2\,.
\end{equation}

\begin{figure}[b]
    \centering
    \includegraphics[width=\linewidth]{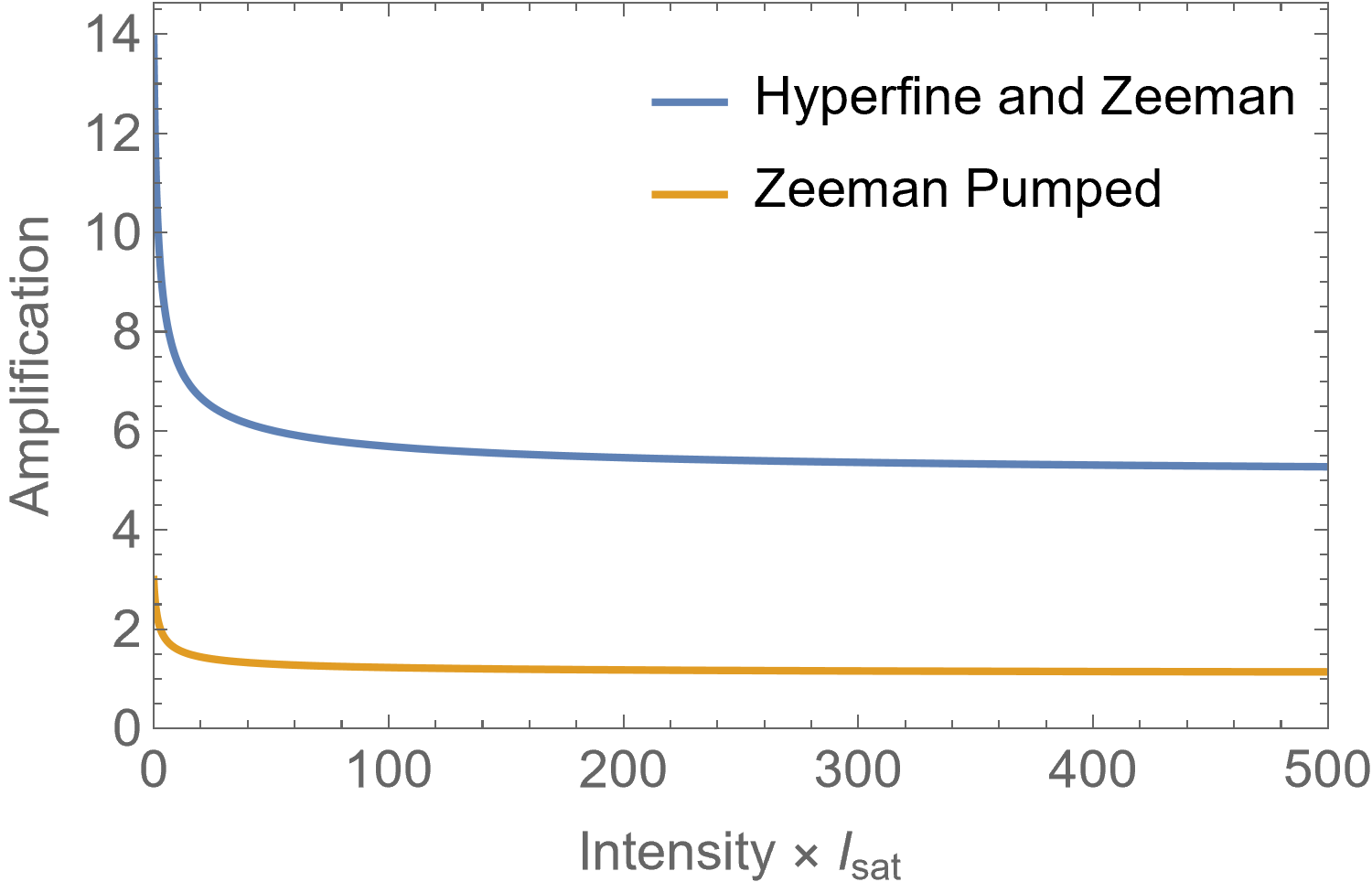}
    \caption{Variation of the amplification factor in $^{163}$Dy with intensity of probe laser beam.}
    \label{fig:ampvintensity}
\end{figure}

The transition amplitude from the Zeeman sublevel $\ket{J, m_J}$ to $\ket{J', m_{J'}}$ due to an E1 transition represented by the operator ($D$) induced by a photon $\ket{k, q}$ at a given electric field strength ($E$) in an atom without nuclear spin is given by
\begin{multline}\label{amplitude Even isotope}
    \braket{J', m_J'|\textbf{D}\cdot \textbf{E}|J, m_J, k, q} = E\times(-1)^{J'-m_J'} \\
    \begin{pmatrix}
        J' & k & J \\
        -m_J' & q & m_J
    \end{pmatrix}
    \braket{J'||D||J}\,,
\end{multline}
and for atoms with nuclear spin\,\cite{auzinsh2010opticallypolarisedatoms, king2008angularmomentumcouplingrabi},
\begin{widetext}
\begin{align}\label{Amplitude odd isotope}
    \braket{F', m_F'|\textbf{D}\cdot \textbf{E}|F, m_F, k, q} = E\times(-1)^{2F'+I+J+1-m_F'} \sqrt{(2F'+1)(1F+1)}
    \begin{Bmatrix}
        J' & 1 & J \\
        F & I & F'
    \end{Bmatrix}
    \begin{pmatrix}
        F' & k & F \\
        -m_F' & q & m_F
    \end{pmatrix}
    \braket{J'||D||J}\,,
\end{align}
\end{widetext}
where the term in the curly brackets are the Wigner-6j symbols and $\braket{J'||D||J}$ is the reduced dipole matrix element. The terms in the parenthesis are the Wigner-3j symbols and are related to the Clebsch-Gordan coefficients ($C$) through
\begin{equation}\label{CG odd}
        \begin{pmatrix}
        F' & k & F \\
        -m_F' & q & m_F
    \end{pmatrix} = (-1)^{F' - m_F - k}\frac{1}{\sqrt{2F + 1}} C_{F, m_F, k, q}^{F', m_F'}.
\end{equation}
We can deduce from here that the saturation parameter can be represented in terms of Clebsch-Gordan coefficients given by,
\begin{equation}
    s = \frac{\braket{F', m_{F'}|\textbf{D}\cdot \textbf{E}|F, m_F}^2}{\Gamma^2} = \big(C_{F, m_F, k, q}^{F', m_F'}\big)^2 \times \frac{I}{I_{sat}}\,,
\end{equation}
where $I$ is the laser intensity and $I_{sat}$ is the saturation intensity of the transition. The amplification factor due to complete optical pumping of the hyperfine states to the $\ket{F=10.5, m_F=10.5}$ hyperfine ground state when excited by the 599\,nm probe is given by,
\begin{equation}\label{Odd isotope amplification}
A = \frac{\rho_{ee}(m_{F'}=9.5) \sum_{m_F, F}\big(C_{F, m_F, 1, 0}^{9.5, 9.5}\big)^2}{\sum_{m_{F'}}\rho_{ee}(m_{F'}) \sum_{m_F, F}\big(C_{F, m_F, 1, 0}^{9.5, m_{F'}}\big)^2}\,,
\end{equation}
where $\rho_{ee}(m)$ is the population of atoms in a particular excited state Zeeman sublevel. An equal distribution of population is assumed in all the ground state Zeeman sublevels when the atoms are not optically pumped. The saturation intensity of the 599\,nm transition is given by, $I_{sat} = hc\pi/2\tau\lambda^3 = 0.11$\,$\mu$W/mm$^2$ and the intensity of the probe laser used in the experiment is $1.27$\,mW/mm$^2$. Fig.\,\ref{fig:ampvintensity} shows the variation in the amplification factor with intensity and the amplification factor at intensities used in the experiment are summarized in Table\,\ref{tab:Calculated amplifications}.

\begin{table}
\begin{ruledtabular}
\begin{tabular}{ccc}
 & \multirow{2}{*}{Zeeman Pumped} & Zeeman and  \\
 & & Hyperfine pumped \\
\hline 
 & \\
$^{162}$Dy and $^{164}$Dy & \multirow{2}{*}{1.2} & \multirow{2}{*}{-} \\
 & \\
$^{163}$Dy and $^{161}$Dy & \multirow{2}{*}{1.2} & \multirow{2}{*}{5.4} \\
($F=10.5 \rightarrow F'=9.5$) &  & \\
&\\
\end{tabular}
\end{ruledtabular}
\caption{\label{tab:Calculated amplifications} Theoretical amplification in the height of the $\sigma_-$ probe spectral lines of the isotopes of Dy.}
\end{table}

\section{Estimation of number of atoms in the polarized states and amplification of the PNC signal}\label{Amplification}

It is evident from Fig.\,\ref{fig:Polarization optimization} and \ref{fig:161 ZeemanandHyperfine} that the atoms in the $F=5.5$ to $F=10.5$ hyperfine ground states are efficiently optically pumped to the $\ket{F=10.5, m_F=10.5}$ ground state. We know sfrom Fig.\,\ref{fig:Zeeman simulation} that any residual unpumped atoms after the 50\,$\mu$s experimental interaction time will exist in the $0<m_F\leq10.5$ Zeeman sublevels. It is also a reasonable assumption to make that any residual atoms in the $0 < m_F < 10.5$ states do not contribute significantly to the amplitude of the probe spectrum due to the reduced density of atoms in these states. With these assumptions, one can estimate the atoms in the $m_F=10.5$ state using Eq.\,\ref{Odd isotope amplification} and the knowledge of the experimental amplification factor ($A$) in the probe spectrum given by,

\begin{multline}\label{AtomsPolarizedState}
\rho_{ee}(m_{F'}=9.5) = \\
A\times\frac{\sum_{m_{F'}}\rho_{ee}(m_{F'}) \sum_{m_F, F}\big(C_{F, m_F, k, q}^{9.5, m_{F'}}\big)^2}{\sum_{m_F, F}\big(C_{F, m_F, k, q}^{9.5, 9.5}\big)^2},
\end{multline}

Assuming an initial unpumped population of one in all the Zeeman sublevels of each hyperfine state, the total number of atoms optically pumped to the $\ket{F=10.5, m_F=10.5}$ state after hyperfine and Zeeman optical pumping with $A=5.9(7)$ is $116$\,($14$). This corresponds to a $\simeq$\,$100$\,\% efficiency in optical pumping. 


\bibliographystyle{ieeetr}
\bibliography{references}

\end{document}